\documentclass[conference,letterpaper]{IEEEtran}
\usepackage[utf8]{inputenc} 
\usepackage[T1]{fontenc}
\usepackage{url}
\usepackage{ifthen}
\usepackage{cite}
\usepackage[cmex10]{amsmath, mathtools} 
\usepackage{amssymb}
\usepackage{xcolor}
\newcommand{\dd}{\mathop{}\!\mathrm{d}}

\begin{document}
\title{Fundamental Limit for One versus Two Point Sources Detection using Direct Imaging} 

\author{%
  \IEEEauthorblockN{Parth Hemant Darekar}
  \IEEEauthorblockA{University of Maryland\\
                    darekar@terpmail.umd.edu}
  \and
  \IEEEauthorblockN{Amit Kumar Jha}
  \IEEEauthorblockA{University of Arizona\\
                    amitjha074@arizona.edu}
  \and
  \IEEEauthorblockN{Michael R. Grace}
  \IEEEauthorblockA{RTX BBN Technologies\\
                    michael.grace@rtx.com}
  \and
  \IEEEauthorblockN{Aqil Sajjad}
  \IEEEauthorblockA{University of Maryland\\
                    asajjad@umd.edu}
  \and
  \IEEEauthorblockN{Saikat Guha}
  \IEEEauthorblockA{University of Maryland\\
                    saikat@umd.edu}  
}

\maketitle

\begin{abstract}
   We consider the task of distinguishing between a single weak incoherent optical point source and two weak incoherent optical point sources located symmetrically about the first source. $\theta$ is the separation between the two point sources scaled to the Point Spread Function (PSF) width in the image plane. Using an ideal focal plane array of intensity detectors (ideal direct imaging), we quantify the performance using the Bhattacharyya distance and find the scaling of its leading order term in terms of $\theta$ in the sub-Rayleigh regime. A suite of previous analyses of this problem lacked a comprehensive analysis for when the amplitude spread function (ASF) of the imaging system has zeros and reported a scaling that we find to be incorrect. We complete this analysis by explicitly calculating the leading order term of the Bhattacharyya distance for ideal direct imaging with any ASF, for small $\theta$ and show the difference in scaling based on the presence or absence of zeros in the ASF. This is similar to the ASF dependent performance in the task of estimating the separation between the two point sources and the task of detecting a change to an object. We then apply our results to the specific example of a Gaussian and a Sinc ASF and show good agreement with numerical calculations. Our results allow the accurate comparison of other measurement schemes with ideal direct imaging, and to the quantum limit.
\end{abstract}

\section{Introduction}
Direct Imaging is the most common method of detecting information-bearing light and consists of a focal plane array (FPA) of intensity detectors. Studying its performance is important in assessing the improvement when using other methods to measure the collected light, and to the fundamental quantum information theoretic limits. Diffraction causes the image plane field arising from a single point source to be in the shape of the ASF---the linear impulse response of the system. The ASF is related to the aperture of the imaging system by a Fourier transform, and the intensity of light on the image plane is given by the squared magnitude of the ASF: known as the PSF. The shot-noise-limited ideal FPA's detection outcome is a spatio-temporal Poisson point process whose rate is this photon-unit intensity impinging the FPA in the image plane. Theoretically, the ASF (and the PSF) may or may not have zeros, a distinction which will be elaborated in more detail later. Most practical ASFs have zeros, e.g., that corresponding to a hard circular aperture function. Therefore, studying the fundamental limits of ideal direct imaging with such ASFs is crucial. When imaging two point sources using direct imaging, the image-plane intensity is a sum of two shifted PSFs. Therefore, if their separation is smaller than the PSF width, it is difficult to resolve them. This is the well known Rayleigh's criterion \cite{rayleigh1879xxxi}. In recent work, Tsang {\em et al.} \cite{tsang2016quantum} showed that the Rayleigh's criterion is {\em not} a fundamental limit for the problem of estimating the separation between two point sources, but a limitation of using focal-plane direct imaging as the receiver. They found that the {\em quantum} information theoretic limits of precision can in principle be achieved by a receiver that separates the image-plane field into a certain set of orthonormal spatial modes prior to (Poisson shot-noise limited) photon detection on those sorted modes.

The work of \cite{krovi2016attaining} considered the problem of distinguishing between one or two incoherently radiating point sources and used the Chernoff exponent \cite{chernoff1952measure ,van2013detection} to compare the performance of different receiver measurements. The Chernoff exponent specifies the rate of decay of the error probability to distinguish between the two hypotheses as the number of collected photons increases. Typically, its scaling with $\theta$ is the metric used for comparison, where $\theta$ is the separation between the two point sources scaled to the PSF width in the image plane. The sub-Rayleigh regime is of significance, where $\theta \ll 1$. However, \cite{krovi2016attaining} only reported numerical results for an ASF with zeros (Sinc ASF) and the leading order term for the Chernoff exponent was not mentioned. In \cite{lu2018quantum}, the leading order term for the Chernoff exponent for ideal direct imaging was reported to be $O\left(\theta^4\right)$. Interestingly, this scaling was reported for all three ASFs considered, with one ASF without zeros (Gaussian) and two ASFs with zeros. The more general problem of distinguishing between arbitrary two dimensional objects was considered in \cite{grace2022identifying}. Here $\theta$ is the $x$- and $y$-dimensional normalized length scales of a general object in a two dimensional image plane. Again, the Chernoff exponent for ideal direct imaging was reported to scale as $O\left(\theta^4\right)$ for both types of ASFs (with or without zeros). An important point to note is that both \cite{lu2018quantum} and \cite{grace2022identifying} did not report numerical results for an ASF with zeros.

We observed from the numerical results of \cite{krovi2016attaining}, which assumed a Sinc-ASF (that has zeros), that the Chernoff exponent of ideal direct imaging scaled as $O\left(\theta^3\right)$---not stated explicitly in \cite{krovi2016attaining}. The main motivation behind our work is to provide a formal proof of the $O\left(\theta^3\right)$ scaling for direct imaging, for {\em any} ASF that has zeros, and to show that the $O\left(\theta^4\right)$ scaling reported in Refs.~\cite{lu2018quantum,grace2022identifying} holds true for ASFs without zeros, e.g., Gaussian ASFs. The quantum Chernoff exponent, achieved by mode sorting receivers, for any ASF is $O\left(\theta^2\right)$.

As for the claim in \cite{lu2018quantum}, we find that their analysis does not work for an ASF with zeros due to division by the ASF in certain Taylor-series terms. The claim in \cite{grace2022identifying} of the $O\left(\theta^4\right)$ scaling for general ASFs was a mere conjecture, made without explicit proof, which we show does not hold for general ASFs, and the Taylor-series technique they used for  their explicit scaling proof assuming a Gaussian ASF does not work for ASFs with zeros. Another motivation here is the observation of the ASF dependent performance of ideal direct imaging for other tasks. For estimating the separation between two point sources, \cite{tsang2016quantum} considered an ASF without zeros and showed a quadratic scaling of the Fisher information with $\theta$ in the sub-Rayleigh regime. However, \cite{paur2019reading} showed analytically that when using an ASF with zeros, this scaling is linear in $\theta$. Furthermore, in the task of detecting a change to an object, a similar observation was made in \cite{grace2023detecting} regarding the scaling of the relative entropy with the Rayleigh-unit object size $\theta$. The scaling was reported to be $O\left(\theta^4\right)$ for an ASF without zeros and numerically observed to be $O\left(\theta^3\right)$ for an ASF with zeros. 

In this work, we consider the specific task of distinguishing between one versus two weak incoherent point sources in the sub-Rayleigh regime ($\theta \ll 1$) for any ASF. We develop a general approach to calculate a closed form expression for the leading order term (in terms of $\theta$) for the Bhattacharyya distance \cite{bhattacharyya1946measure, kailath1967divergence} of ideal direct imaging. In general, the Bhattacharyya distance is a special case of the Chernoff exponent \cite{kailath1967divergence} and is a lower bound to it. Both quantities coincide when the minimization required for the Chernoff exponent occurs at a specific value (at $s = 1/2$). Two different calculations with distinct methods are needed: one for an ASF without zeros and one for an ASF with zeros. The former is similar to the approach in \cite{lu2018quantum}. Our results prove the scaling of $O\left(\theta^4\right)$ for any ASF without zeros, which matches the results in \cite{lu2018quantum,grace2022identifying}. More importantly, we uncover the scaling of $O\left(\theta^3\right)$ for any ASF with zeros. This was previously unreported and was incorrectly claimed to be $O\left(\theta^4\right)$. We apply our results to two examples: a Gaussian ASF (ASF without zeros) and a Sinc ASF (ASF with zeros), and show good agreement with numerical calculations.

\section{Problem Setup}
\begin{figure}[htbp]
   \centering
   \includegraphics[width=0.35\textwidth]{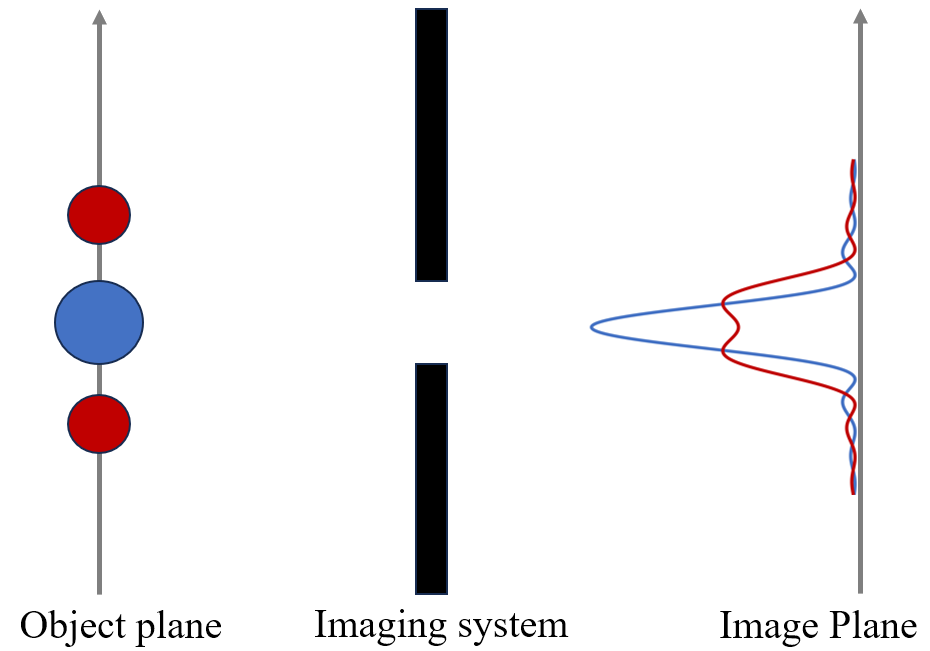}
   \caption{Schematic of the setup. Under $H_1$, we have a weak thermal source (in blue) at the origin. Under $H_2$, we have two equally bright, weak thermal sources (in red) with an image plane separation of $\theta$ that is scaled to the PSF width. For small separation, the intensity distributions of light on the image plane under each hypothesis become difficult to distinguish.}
   \label{fig:schematic}
 \end{figure}
The problem setup is shown in Fig. \ref{fig:schematic}. A weak thermal source (in blue) is imaged at the origin (hypothesis $H_1$). Two equally bright weak thermal sources (hypothesis $H_2$, in red) are imaged at positions $-\theta/2$ and $\theta/2$. We are interested in distinguishing between the two hypotheses using ideal direct imaging. The analysis is for a single dimension and the total brightness is the same for each hypothesis with mean photon number $\varepsilon$ per temporal mode. For weak sources, relevant to optical-frequency astronomical imaging for example, we have $\varepsilon \ll 1$. We focus our analysis on a single photon in each temporal mode of the collected radiation field. For ideal direct imaging, the probabilities of detecting a photon at position $x$ under hypothesis $H_1$ or $H_2$ are: 
\begin{IEEEeqnarray}{rCl}
    P(x|H_1) & = & \left|\psi(x)\right|^2, \\
    P(x, \theta|H_2)& = & \frac{1}{2} \left( \left|\psi(x + \theta/2)\right|^2 +\left|\psi(x - \theta/2)\right|^2 \right).
\end{IEEEeqnarray}
Here $\psi(x)$ is the ASF. For ease of notation, we label $p_1(x) = P(x|H_1)$ and $p_2(x, \theta) = P(x, \theta|H_2)$. Then the Bhattacharyya distance $\xi^{(1)}_{\text{direct}}(\theta)$ \cite{kailath1967divergence} can be written as  
\begin{IEEEeqnarray}{rCl}
    \xi^{(1)}_{\text{direct}}(\theta) & = & -\log\left[B(\theta) \right], \label{eq:bha} \\
    B(\theta) & = & \int_{-\infty}^\infty \sqrt{p_1(x) p_2(x, \theta)}\dd x. \label{eq:qs} 
\end{IEEEeqnarray}
Here, the superscript indicates the conditioning on a photon arriving on a scene, similar to \cite{lu2018quantum}. We now find an alternate expression for $\xi^{(1)}_{\text{direct}}(\theta)$ as follows. Consider the following term \begin{IEEEeqnarray}{rCl}
    \sqrt{p_1(x) p_2(x, \theta)} & = & \frac{1}{2} p_1(x) + \frac{1}{2} p_2(x, \theta) \nonumber \\ 
    & & \negmedspace{} - \frac{1}{2} \left[\sqrt{p_2(x, \theta)} - \sqrt{p_1(x)} \right]^2 \nonumber.
\end{IEEEeqnarray}
Integrating over $x$ gives us
\begin{IEEEeqnarray}{rCl}
    B(\theta) & = & \left[ 1 - \frac{1}{2} I(\theta) \right], \label{eq:qs_modified}
\end{IEEEeqnarray}
where 
\begin{IEEEeqnarray}{rCl}
    I(\theta) & = & \int_{-\infty}^{\infty} \left[\sqrt{p_2(x, \theta)} - \sqrt{p_1(x)} \right]^2 \dd x. \label{eq:integrand}
\end{IEEEeqnarray}
To obtain (\ref{eq:qs_modified}), we used (\ref{eq:qs}) and the property that $p_1(x)$ and $p_2(x, \theta)$ integrate to $1$. Our aim is to simplify $I(\theta)$ for $\theta \ll 1$ and find the leading order term for $\xi^{(1)}_{\text{direct}}(\theta)$. In the next section, we analyze the case when the ASF $\psi(x)$ is without zeros. The reason for conditioning the analysis on the existence of the zeros of $\psi(x)$ will become clear at the end of the next section. The probability of classification error, $P_e^{(N)} \sim e^{-N\,\xi^{(1)}_{\text{direct}}(\theta)}$, where $N$ is the total number of photons collected during the collection window.

\section{Amplitude Spread Function without zeros}
When $\psi(x)$ does not have zeros, $p_1(x) = \left|\psi(x)\right|^2$ is bounded away from $0$. We write a Maclaurin series expansion for $\sqrt{p_2(x, \theta)}$ in terms of $\theta$ as 
\begin{IEEEeqnarray}{rCl}
    \sqrt{p_2(x, \theta)} & = & \sqrt{p_2(x, \theta)} \big|_{\theta=0} + \left. \frac{\partial \sqrt{p_2(x, \theta)}}{\partial \theta}\right|_{\theta=0} \frac{\theta^1}{1!} \nonumber \\
    & & \negmedspace{} + \left. \frac{\partial^2 \sqrt{p_2(x, \theta)}}{\partial \theta^2}\right|_{\theta=0} \frac{\theta^2}{2!} + O(\theta^3). \label{eq:maclaurin}
\end{IEEEeqnarray}
Calculating each term and rearranging lets us write
\begin{IEEEeqnarray}{rCl}
    \sqrt{p_2(x, \theta)} - \sqrt{p_1(x)} & = & \frac{1}{16 \left|\psi(x)\right|} \left[\frac{d^2 \big(\left|\psi(x) \right|^2 \big)}{dx^2} \right] \theta^2 \nonumber \\ 
    & & \negmedspace{} + O(\theta^3). 
\end{IEEEeqnarray}
Substituting in (\ref{eq:integrand}) gives
\begin{IEEEeqnarray}{rCl}
    I(\theta) & = & \frac{\theta^4}{256} \int_{-\infty}^{\infty} \left(f(x)\right)^2 \dd x + O(\theta^5). \label{eq:integrand_without_zeros} 
\end{IEEEeqnarray}
where $f(x)$ is defined as
\begin{IEEEeqnarray}{rCl}
    f(x) = \frac{1}{\left|\psi(x)\right|} \frac{d^2 \big(\left|\psi(x) \right|^2 \big)}{dx^2}. \label{eq:fx}
\end{IEEEeqnarray}
Substituting this expression in (\ref{eq:qs_modified}), we get 
\begin{IEEEeqnarray}{rCl}
   B(\theta) = 1 - \frac{\theta^4}{512} \int_{-\infty}^{\infty} \big(f(x)\big)^2 \dd x + O(\theta^5).
\end{IEEEeqnarray}
Therefore, the Bhattacharyya distance for $\theta \ll 1$ is 
\begin{IEEEeqnarray}{rCl}
    \xi^{(1)}_{\text{direct}}(\theta) = \frac{\theta^4}{512} \int_{-\infty}^\infty \big(f(x)\big)^2 \dd x + O(\theta^5). \label{eq:bha_without_zeros}
\end{IEEEeqnarray}
Here, we used the series expansion for $\log(\cdot)$ and $f(x)$ is defined in (\ref{eq:fx}). The leading order term in the sub-Rayleigh regime is $O\left(\theta^4 \right)$, matching the result in \cite{lu2018quantum}, in which the calculated Chernoff exponent coincides with the Bhattacharyya distance. An important observation is that $\psi(x)$ is in the denominator in $f(x)$. This is fine when $\psi(x)$ does not have any zeros, but when there are zeros the integral diverges. Therefore, we cannot use this analysis for an ASF with zeros. We need a different approach as shown in the next section.

\section{Amplitude Spread Function with zeros}
When $\psi(x)$ has zeros, the integral in (\ref{eq:bha_without_zeros}) diverges. Hence, we cannot use the Maclaurin series expansion of (\ref{eq:maclaurin}). Let $\psi(x)$ have simple isolated zeros at $x=x_n$, where $n$ is a label. Taking inspiration from asymptotic methods, we divide the region of integration for $I(\theta)$ in (\ref{eq:integrand}) into parts - near the zeros and away from the zeros. Near a zero at $x_n$ we define an interval $R_n = \{x \mid |x-x_n| \leq \delta\}$, where $\delta > 0$ is fixed to be small. Therefore, the region of integration near the zeros is $R=\cup_n R_n$. The region of integration away from the zeros is then $R^c$. This allows us to write 
\begin{IEEEeqnarray}{rCl}
    I(\theta) & = & I_{\text{near}}(\theta) + I_{\text{away}}(\theta),
\end{IEEEeqnarray} 
where $I_{\text{near}}(\theta)$ and $I_{\text{away}}(\theta)$ are defined as 
\begin{IEEEeqnarray}{rCl}
    I_{\text{near}}(\theta) & = & \sum_{n} \int_{R_n} \left[\sqrt{p_2(x, \theta)} - \sqrt{p_1(x)} \right]^2 \dd x, \\
    I_{\text{away}}(\theta) & = & \int_{R^c} \left[\sqrt{p_2(x, \theta)} - \sqrt{p_1(x)} \right]^2 \dd x .
\end{IEEEeqnarray}
Away from the zeros, we can use the previous analysis for an ASF without zeros by changing the region of integration in (\ref{eq:integrand_without_zeros}) to get
\begin{IEEEeqnarray}{rCl}
    I_{\text{away}}(\theta) & = & \frac{\theta^4}{256}  \int_{R^c} \left(f(x)\right)^2 \dd x + O(\theta^5),
\end{IEEEeqnarray}
where $f(x)$ was defined in (\ref{eq:fx}). 

To calculate $I_{\text{near}}(\theta)$, we write a Taylor series expansion for $\psi(x)$ about a zero at $x = x_n$ as
\begin{IEEEeqnarray}{rCl}
    \psi(x) & = & \left.\psi(x)\right|_{x=x_n} + \left.\frac{d\psi(x)}{dx}\right|_{x=x_n} \frac{(x-x_n)^1}{1!} \nonumber \\
    & & \negmedspace{} + O\big((x-x_n)^2\big), \nonumber \\
    & = & g(x_n) t + O\left(t^2\right).
\end{IEEEeqnarray}
Here we have substituted $t = x - x_n$, $|t|\leq\delta$ and $g(x_n)$ is defined as 
\begin{IEEEeqnarray}{rCl}
    g(x_n) & = & \left. \frac{d\psi(x)}{dx}\right|_{x=x_n}. \label{eq:gxn}
\end{IEEEeqnarray}
Using the fact that $\theta \ll 1$, we write $\psi(x\pm \theta/2) = g(x_n) (t \pm \theta/2) + O\big((t\pm \theta/2)^2 \big)$. Then the leading order term for $p_1(x)$ and $p_2(x, \theta)$ can be written as
\begin{IEEEeqnarray}{rCl}
    p_1(x) & \approx & \left|g(x_n)\right|^2 t^2, \\
    p_2(x) & \approx & \left|g(x_n)\right|^2 \left(t^2 + \theta^2/4 \right). 
\end{IEEEeqnarray}
Using the above,
\begin{IEEEeqnarray}{rCl}
    \sqrt{p_2(x, \theta)} - \sqrt{p_1(x)} & \approx & \left|g(x_n)\right|  \left[ \sqrt{t^2 + \theta^2/4} - |t|\right].
\end{IEEEeqnarray}
Substituting into the expression for $I_{\text{near}}(\theta)$,
\begin{IEEEeqnarray}{rCl}
    I_{\text{near}}(\theta) & \approx & \sum_{n} \left|g(x_n)\right|^2 \int_{-\delta}^\delta  \Bigg[ 2t^2 -  2|t| \sqrt{t^2 + \theta^2/4}  \nonumber \\
    & & \negmedspace{}  + \frac{\theta^2}{4}  \Bigg] \dd t, \nonumber \\
    & = & 2 \sum_{n} \left|g(x_n)\right|^2 \Bigg[ \frac{2\delta^3}{3}  - \frac{2}{3} \left(\delta^2 + \theta^2/4 \right)^{3/2} \nonumber \\
    &  & \negmedspace{}  + \frac{\theta^3}{12} + \frac{\delta \theta^2}{4} \Bigg].  
\end{IEEEeqnarray}
The last line used the fact that the integrand is even. Since we are interested in the regime where $\theta \ll 1$, we calculate an asymptotic expression for $\left(\delta^2 + \theta^2/4 \right)^{3/2}$ when $\theta \rightarrow 0^+$ as 
\begin{IEEEeqnarray}{rCl}
    \frac{2}{3}\left(\delta^2 + \theta^2/4 \right)^{3/2} & = & \frac{2\delta^3}{3} + \frac{\delta \theta^2}{4} + \frac{\theta^4}{64\delta} + O\left(\theta^6\right).
\end{IEEEeqnarray}
Observe that the $O\left(\theta^4\right)$ term has $\delta$ in the denominator. But we had chosen $\delta>0$ whereas $\theta \rightarrow 0^+$. Specifically, we require that $\theta^4/\delta \ll \theta^3$. This allows us to ignore the $O\left(\theta^4\right)$ term as compared to the $O\left(\theta^3\right)$ term as $\theta \rightarrow 0^+$. Therefore, we can write $I_{\text{near}}(\theta)$ as 
\begin{IEEEeqnarray}{rCl}
    I_{\text{near}}(\theta) & \approx & \frac{1}{6} \sum_{n} \left|g(x_n)\right|^2 \theta^3.
\end{IEEEeqnarray}
Interestingly, the $O\left(\delta\right)$ and $O\left(\delta^3\right)$ terms get canceled and the leading order term is independent of $\delta$. The above expression is also the leading order term for $I(\theta)$ for $\theta \ll 1$ because the leading order term for $I_{\text{away}}(\theta)$ is $O\left(\theta^4 \right)$. Substituting this into (\ref{eq:qs_modified}), we get
\begin{IEEEeqnarray}{rCl}
    B(\theta) & \approx & 1 - \frac{\theta^3}{12} \sum_{n} \left|g(x_n)\right|^2.
\end{IEEEeqnarray}
Therefore, the Bhattacharyya distance for $\theta \ll 1$ is 
\begin{IEEEeqnarray}{rCl}
    \xi^{(1)}_{\text{direct}}(\theta) & \approx  &  \frac{\theta^3}{12} \sum_{n} \left|g(x_n)\right|^2. \label{eq:bha_with_zeros}
\end{IEEEeqnarray}
Here, we used the series expansion for $\log(\cdot)$ and $g(x_n)$ is defined in (\ref{eq:gxn}). Recall that we never specified the value of $\delta$. But remarkably in the sub-Rayleigh regime ($\theta \ll 1$), the leading order term for $\xi^{(1)}_{\text{direct}}(\theta)$ does not depend on $\delta$. Therefore, we limit our definition of $\delta$ to a small fixed quantity greater than $0$ such that $\theta^4/\delta \ll \theta^3$. Comparing (\ref{eq:bha_without_zeros}) with (\ref{eq:bha_with_zeros}) shows that the Bhattacharyya distance for an ASF with zeros is better than the Bhattacharyya distance for an ASF without zeros by an order of magnitude. We now apply our results to the specific example of a Gaussian and a Sinc ASF. We also compare these results to numerical calculations.

\section{Examples}
We first consider the example of a Gaussian ASF (ASF without zeros) and use (\ref{eq:bha_without_zeros}) to calculate the leading order term of the Bhattacharyya distance when $\theta \ll 1$. We compare this result to the results obtained by numerically evaluating (\ref{eq:bha}). We then repeat this analysis for a  Sinc ASF (ASF with zeros) by using (\ref{eq:bha_with_zeros}). 

\subsection{Gaussian Amplitude Spread Function}
\begin{figure}[htbp]
   \centering
   \includegraphics[width=0.45\textwidth]{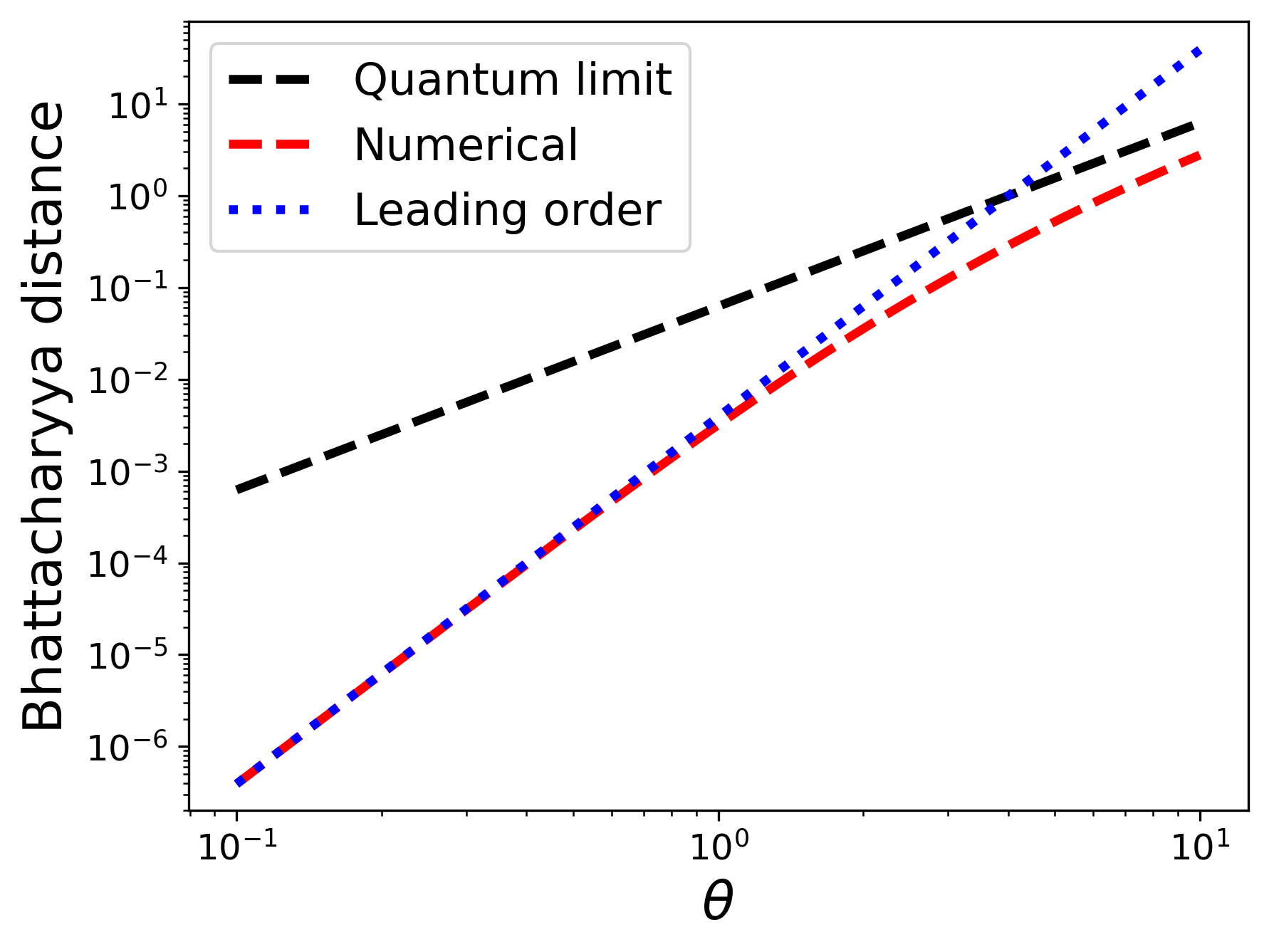}
   \caption{Bhattacharyya distance for ideal direct imaging for the Gaussian Amplitude Spread Function (ASF) as a function of $\theta$. The numerical results (red, dashed) show good agreement with the leading order term (blue, dotted) when the separation $\theta$ is small. The Bhattacharyya distance scales as $O\left(\theta^4\right)$ in the sub-Rayleigh regime ($\theta \ll 1$). The black dashed line is the quantum Chernoff exponent that scales as $O\left(\theta^2\right)$.}
   \label{fig:gaussian}
 \end{figure}

Consider $\psi(x) = \left(2 \pi \right)^{-1/4} \exp\left(-x^2/4\right)$.
Using (\ref{eq:fx}), 
\begin{IEEEeqnarray}{rCl}
    f(x) & = & \left({2} \pi \right)^{-1/4} \left(x^2 - 1\right) \exp\left(-x^2/4\right), \nonumber \\
    \int_{-\infty}^\infty \big(f(x)\big)^2 \dd x & = & {2}. \nonumber
\end{IEEEeqnarray}
Substituting into (\ref{eq:bha_without_zeros}) gives the leading order term as 
\begin{IEEEeqnarray}{rCl}
    \xi^{(1)}_{\text{direct, Gaussian}}(\theta) & \approx & \frac{\theta^4}{256}.
\end{IEEEeqnarray}
This matches the result calculated in \cite{lu2018quantum}. The quantum Chernoff exponent is calculated using \cite{kargin2005chernoff} as $\xi^{(1)}_{\text{Q, Gaussian}}(\theta) = \theta^2/16$. In Fig. \ref{fig:gaussian}, we plot the leading order term (blue, dotted) of $\xi^{(1)}_{\text{direct, Gaussian}}(\theta)$ calculated above and compare it to the numerical calculations (red, dashed). We see a good agreement when $\theta$ is small. Compared to $\xi^{(1)}_{\text{Q, Gaussian}}(\theta)$, we observe a quadratic gap in the scaling in terms of $\theta$.

\subsection{Sinc Amplitude Spread Function}
\begin{figure}[htbp]
   \centering
   \includegraphics[width=0.45\textwidth]{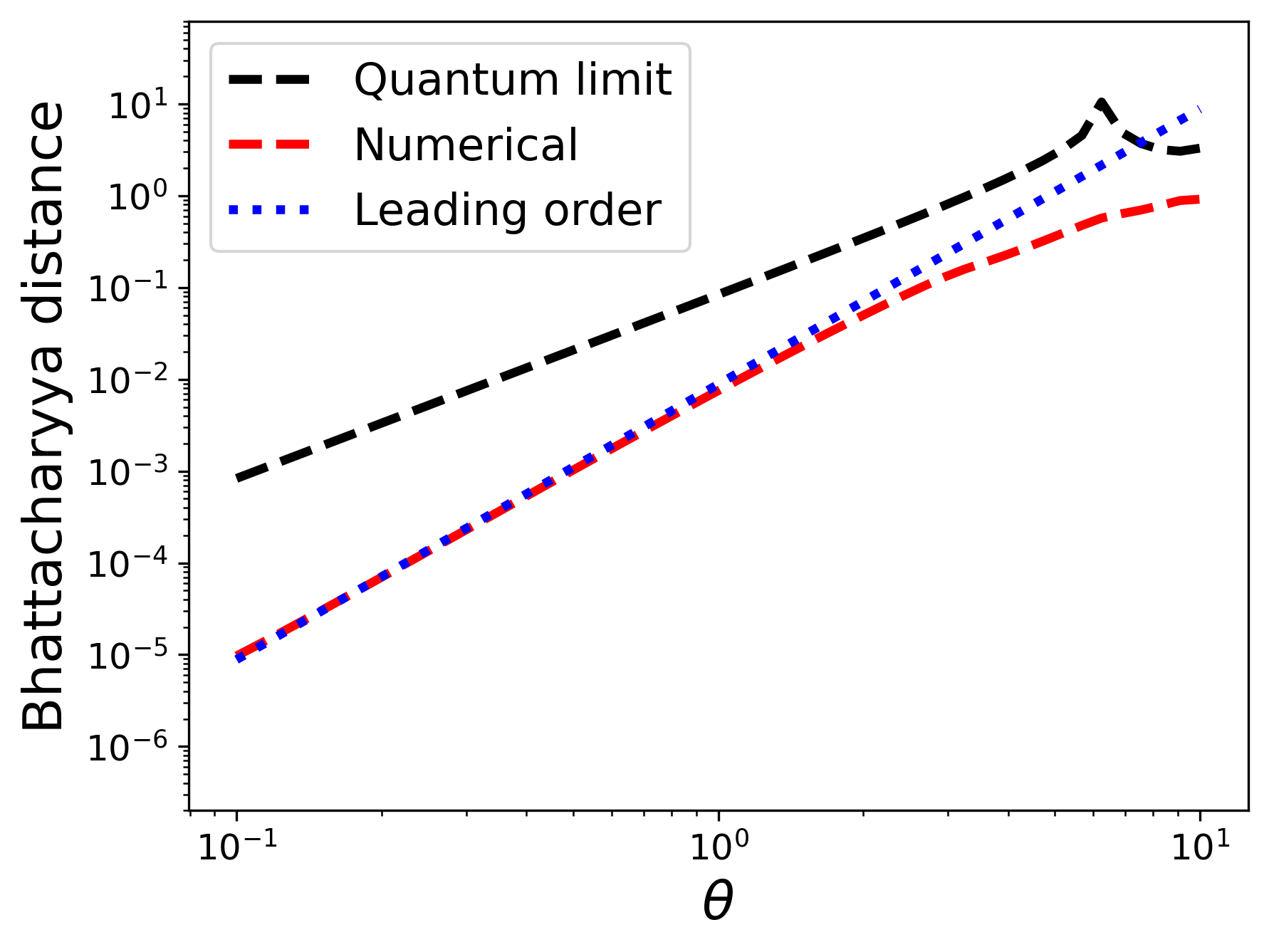}
   \caption{Bhattacharyya distance for ideal direct imaging for the Sinc Amplitude Spread Function (ASF) as a function of $\theta$. The numerical results (red, dashed) show good agreement with the leading order term (blue, dotted) when the separation $\theta$ is small. The Bhattacharyya distance scales as $O\left(\theta^3\right)$ in the sub-Rayleigh regime ($\theta \ll 1$). The black dashed line is the quantum Chernoff exponent that scales as $O\left(\theta^2\right)$.}
   \label{fig:sinc}
 \end{figure}

Consider $\psi(x) = (1/\sqrt{\pi})\text{sinc}\left(x\right)$, where $\text{sinc}(x) \coloneqq \sin(x)/x$. The zeros of $\psi(x)$ are at $x = x_n = n\pi$, $n \in \mathbb{Z} - \{0\}$. In this notation, the summation over all the zeros will be a summation over $n, n \in \mathbb{Z} - \{0\}$. Using (\ref{eq:gxn}), we get $g(x_n) = (-1)^n/\left(n\pi^{3/2}\right)$ and 
\begin{IEEEeqnarray}{rCl}
    \sum_{n \in \mathbb{Z} - \{0\}} \left|g(x_n)\right|^2 & = &  \frac{1}{\pi^3} \sum_{n \in \mathbb{Z} - \{0\}} \frac{1}{n^2} = \frac{1}{3\pi}. \nonumber 
\end{IEEEeqnarray}
Substituting into (\ref{eq:bha_with_zeros}) gives the leading order term as 
\begin{IEEEeqnarray}{rCl}
    \xi^{(1)}_{\text{direct, Sinc}}(\theta) & \approx & \frac{\theta^3}{36\pi}.
\end{IEEEeqnarray}
The quantum Chernoff exponent is calculated using \cite{kargin2005chernoff} as $\xi^{(1)}_{\text{Q, Sinc}}(\theta) = \theta^2/12 + O\left(\theta^4\right)$. In Fig. \ref{fig:sinc}, we plot the leading order term (blue, dotted) of $\xi^{(1)}_{\text{direct, Sinc}}(\theta)$ calculated above and compare it to the numerical calculations (red, dashed). We see good agreement when $\theta$ is small. Compared to $\xi^{(1)}_{\text{Q, Sinc}}(\theta)$, we observe a linear gap in the scaling in terms of $\theta$.

\section{Conclusion}
In this paper, we have considered the task of distinguishing between one and two weak incoherent point sources using ideal direct imaging. We have shown that the leading order term of the Bhattacharyya distance in the sub-Rayleigh regime depends on the existence of zeros of the amplitude spread function (ASF) of the imaging system. For an ASF without zeros we have shown a scaling of $O\left(\theta^4\right)$ for $\theta \ll 1$, where $\theta$ is the separation between the two point sources scaled to the Point Spread Function (PSF) width in the image plane. This result is in agreement with prior work, where the Bhattacharyya distance coincides with the Chernoff exponent. For an ASF with zeros, we have shown a scaling of $O\left(\theta^3\right)$ that was previously unreported and claimed to be $O(\theta^4)$. We have also shown good agreement with numerical calculations for a Gaussian ASF and a Sinc ASF. Since most practical systems have ASFs with zeros, our results provide an accurate reference to compare alternative measurement schemes, such as modal imaging receivers~\cite{grace2022identifying}, with ideal direct imaging. The quantum optimal scaling of the Chernoff exponent for any ASF is $O\left(\theta^2\right)$. We were also able to obtain the $O\left(\theta^3\right)$ scaling for the Chernoff exponent for an ASF with zeros. However, that analysis is not included here because of additional work required to match the theoretical optimal value of $s$ with the numerical result. An immediate extension of this work is to the task of discriminating between arbitrary objects in two dimensions and to detect a change to existing objects using an ASF with zeros, both of which will be the subject of a future paper. Finally, we note that the ASF dependent performance of ideal direct imaging is similar to the observations in other imaging tasks of estimating the separation between two point sources and detecting a change to an object. In fact, a simple application of our result to an argument in Section 3.2 of~\cite{Grace2022thesis}, shows that the direct-imaging Fisher Information for the relative error in estimating the two-point source separation, when the imaging-system ASF has zeros, scales as $O\left(\theta^3\right)$.

\section*{Acknowledgment}
We acknowledge helpful conversations with Nico Deshler and Vincenzo Tamma; and ARO MURI grant W911NF2110325 and AFOSR grant FA9550-24-1-0139, for funding support. 
\nocite{submission_status}
\enlargethispage{-0.9in}

\bibliographystyle{IEEEtran}
\bibliography{references}

\end{document}